\documentclass[conference]{IEEEtran}
\IEEEoverridecommandlockouts
\usepackage{cite}
\usepackage{amsmath,amssymb,amsfonts}
\usepackage{algpseudocode}
\usepackage{algorithm}
\usepackage{hyperref}
\usepackage{pgfplots}
\usepackage{multirow}
\usepackage{soul}
\usepackage{tcolorbox}

\usepackage{enumitem}

\usepackage{graphicx}
\usepackage{textcomp}
\usepackage{xcolor}
\def\BibTeX{{\rm B\kern-.05em{\sc i\kern-.025em b}\kern-.08em
    T\kern-.1667em\lower.7ex\hbox{E}\kern-.125emX}}
\begin{document}

\title{Explainability Guided Adversarial Evasion Attacks on Malware Detectors
}

\author{\IEEEauthorblockN{Kshitiz Aryal and Maanak Gupta}
\IEEEauthorblockA{\textit{Department of Computer Science} \\
\textit{Tennessee Tech University}\\
Cookeville, USA\\
karyal42@tntech.edu, mgupta@tntech.edu}
\and
\IEEEauthorblockN{Mahmoud Abdelsalam}
\IEEEauthorblockA{\textit{Department of Computer Science} \\
\textit{North Carolina A\&T State University}\\
Greensboro, USA \\
mabdelsalam1@ncat.edu}
\and
\IEEEauthorblockN{Moustafa Saleh}
\IEEEauthorblockA{\textit{Oracle Cloud Infrastructure} \\
\textit{Oracle}\\
Seattle, USA \\
moustafa.saleh@oracle.com}
}

\maketitle

\begin{abstract}
As the focus on security of Artificial Intelligence (AI) is becoming paramount, research on crafting and inserting optimal adversarial perturbations has become increasingly critical. In the malware domain, this adversarial sample generation relies heavily on the accuracy and placement of crafted perturbation with a goal to evade a trained classifier. This work focuses on applying explainability techniques to enhance the adversarial evasion attack on a machine-learning-based Windows PE malware detector. The explainable tool identifies the regions of PE malware files that have the most significant impact on the decision-making process of a given malware detector, and therefore, the same regions can be leveraged to inject the adversarial perturbation for maximum efficiency. Profiling all the PE malware file regions based on their impact on the malware detector's decision enables the derivation of an efficient strategy for identifying the optimal location for perturbation injection. The strategy should incorporate the region's significance in influencing the malware detector's decision and the sensitivity of the PE malware file's integrity towards modifying that region.

To assess the utility of explainable AI in crafting an adversarial sample of Windows PE malware, we utilize \textit{DeepExplainer} module of SHAP (SHapley Additive exPlanations) for determining the contribution of each region of PE malware to its detection by a CNN-based malware detector, MalConv. The analysis includes both local and global explanations for the given malware samples. We performed the functionality-preserving adversarial perturbation injection in different regions of PE malware wherever possible while performing non-functionality-preserving operations in a few remaining regions. This approach allows us to examine the relationship between SHAP values and the evasion rate of the adversarial attack. Furthermore, we analyzed the significance of SHAP values at a more granular level by subdividing each section of Windows PE into small subsections. We then performed an adversarial evasion attack on the subsections based on the corresponding SHAP values of the byte sequences. Our experimental evaluation shows a significant improvement in the success and efficiency of adversarial evasion attacks when injecting the perturbation in PE malware locations based on SHAP values compared to random PE locations.

\end{abstract}


\section{Introduction}
Windows PE malware binary has been a prime target for Adversarial Evasion (AE) attacks in the malware domain, and the reason can be attributed to the abundance or the discrepancies of PE file formats~\cite{nisi2021lost,aryal2021survey}. The AE attack on a malware detector is carried out by modifying a malicious binary file in a way that it gets recognized as a benign file by a target malware detector. To formally define an AE attack, let's consider a machine-learning-based malware detector $C$ and a PE malware binary $B$ such that $C(B) = Y_{Mal}$, where $Y_{Mal}$ is the real label of $B$. Now, the adversarial evasion attack introduces the perturbation $\delta$ to a malware binary $B$ in such a way that $C(B+\delta) = Y_{Ben}$, where $Y_{Ben}$ is a new adversarial label of $B$ and $Y_{Mal} \neq Y_{Ben}$. A major challenge in carrying out AE attacks in the malware domain arises due to the strict semantic constraints of binary executable. Unlike other domains, a single random perturbation in a binary executable can break its executability and functionality~\cite{aryal2022analysis}. Therefore, a successful AE attack on a malware domain should generate the perturbations that not only evades the target malware detector but also preserves the malware file's integrity and behaviour. 

Given the absolute requirement of preserving the malware's functionality and executability during AE attacks, early research\cite{kolosnjaji2018adversarial, kreuk2018deceiving} in this domain concentrated on crafting the evasive malware with the utmost caution, refraining from altering the structure of malware. Many studies~\cite{hu2018black, rosenberg2020generating} confined their efforts to craft adversarial evasion only on feature space (feature level) rather than shifting to the problem space, thereby limiting the practical applicability of their work. Here, The problem space refers to the PE file where it exists in its standard defined format, whereas the feature space can be static or dynamic features obtained from some malware analysis techniques.
Recent works explored the feasibility of injecting the adversarial perturbations across different regions within Windows PE malware, including the header~\cite{demetrio2019explaining, demetrio2021adversarial}, slack spaces~\cite{suciu2019exploring}, codecaves~\cite{yuste2022optimization} and other regions while keeping the malware file intact. Most lately, a novel approach proposed by Aryal et al. ~\cite{aryal2024intrasection} inserted the adversarial perturbations within different sections of PE malware files while preserving the integrity of malware. This approach injects code caves within any sections of the PE malware file to make space for adversarial perturbations, without altering the malware's integrity. As the various locations to inject perturbation evolve across the PE malware structure, the research focus is transitioning towards optimizing the efficiency of injected perturbation. The goal is to craft a functionality-preserved evasive adversarial malware with minimal perturbation injection.

A promising approach to increase the efficiency of adversarial perturbation being injected in Windows PE malware is through the utilization of \textit{explainable AI}. Explainability techniques are inherently employed for understanding and interpreting the models, leading to enhanced trustworthiness and transparency of black-box machine learning models~\cite{vigano2020explainable, nadeem2023sok,manthena2023analyzing}.  
In a prior study, Rosenberg et al.~\cite{rosenberg2020generating} used explainability to perform adversarial attacks on feature space and were limited due to the practicality issues of mapping feature space modifications to problem space. Another work by Demetrio et al.~\cite{demetrio2019explaining} used integrated gradients for feature attribution, discovering the high efficiency of perturbation in the PE header regions. However, the study lacked comprehensive analysis on utilizing the explainability of a model for enhancing adversarial evasion attacks. Their analysis is primarily confined to the PE header region, overlooking the potential impact of explainability in other regions of the PE malware. Furthermore, this work could not address the different levels of granularity that can be used to inject perturbations within the PE regions with the use of explainability. 
To address these limitations, we propose an \textit{"Explainability Guided Adversarial Evasion Attacks on Malware Detectors"}. 

In this paper, we explore the potential of explainability techniques to strategize efficient AE attacks by leveraging the reasoning behind a  machine learning's decision on the Windows PE malware detector. 
For this study, we target a widely used CNN-based malware detector, MalConv~\cite{raff2018malware}, as a malware detector trained on Windows PE malware and benign files. Our approach involves calculating the SHAP values for each byte of a PE binary malware and aggregating these SHAP values corresponding to different PE regions to determine the collective contribution of individual PE regions. We assess the impact of perturbation in different regions coupling to the corresponding SHAP values. Additionally, we divide larger Windows PE malware sections into smaller subsections and aggregate the respective SHAP values to enable more precise control over the placement of adversarial perturbation. 
The major \textbf{contributions} of this work are listed below:
\begin{itemize}[leftmargin=*]
    \item Explain the contributions of each byte of Windows PE malware binary towards its detection by a CNN model using SHAP values and map it to the regions of the PE structure;
    \item Evaluate and compare local and aggregated global SHAP values of different regions in the PE malware structure;
    \item Perform adversarial evasion attack by injecting perturbations in regions of the PE malware and correlate the results with the corresponding SHAP values;
    \item Divide larger PE malware sections into smaller subsections and compute their corresponding SHAP values to evaluate an adversarial injection strategy at a more fine-grained level.
\end{itemize}

The rest of the paper is organized as follows. 
Section \ref{sec:related} presents the background information on PE structure, malware detection, adversarial malware and explainability.
Section \ref{sec:exp-ae} discusses our proposed methodology for computing and analyzing SHAP values, leading to adversarial evasion attacks. 
Section \ref{sec:experiment} discuss experimental evaluation and analysis of the results, followed by summary in Section \ref{sec:summary}.

\section{Background}
\label{sec:related}
\subsection{Windows PE Structure}
\begin{figure}
    \centering
    \includegraphics[scale=0.8]{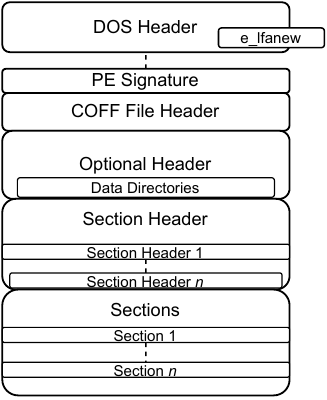}
    \caption{Structure of Windows PE Malware}
    \vspace{-5mm}
    \label{fig:structure_of_pe}
\end{figure}
The understanding of Windows Portable Executable (PE) structure plays a crucial role in adversarial malware not only because of the idiosyncratic nature of different parts but also because of differences in sensitivity to the modification. The Windows PE format is a standard format supported by the Windows operating systems~\cite{PEFormat69:online}. The structure of the PE file, as shown in Figure \ref{fig:structure_of_pe}, is composed of linear data streams divided into different regions. The structure starts with the \textit{DOS header}, which was initially used in the DOS operating system used for backward compatibility. The following 4-byte \textit{ PE signature} identifies the file as a PE format image file. The \textit{COFF header} contains information like supported architecture, size of section table, and characteristics flags denoting the attributes of a file. The \textit{Optional Header} holds details about image files and contains important fields like \textit{MajorOperatingSystemVersion}, \textit{MinorOperatingSystemVersion}, \textit{SizeOfCode}, \textit{AddressOfEntryPoint}, \textit{BaseOfCode}, \textit{SizeofStackCommit}, \textit{SizeOfHeapCommit} etc. The end portion of \textit{Optional Header} has \textit{Data Directories}, which holds a relative virtual address and the size of a table or string used by Windows. The \textit{Section Header} contains details of individual sections, including its \textit{Name}, \textit{VirtualSize}, \textit{VirtualAddress}, size, pointers and characteristics flags. Finally, the sections do the job of organizing the executable files logically. Each section has its own purpose, like \texttt{.text} section contains executable code, \texttt{.data} contains initialized code, etc. To be consistent, in this paper we use ``PE region" to refer to the regions in the PE file such as \textit{DOS header}  \textit{COFF header},  \textit{Optional Header} etc., and the term ``Section" refers to the PE sections such as \texttt{.text}, \texttt{.data}, \texttt{.rdata} etc. The term ``Location" refers to the position of malware bytes within a PE file.

\subsection{Malware Detection}
\begin{figure}[!t]
    \centering
    \includegraphics[scale=0.6]{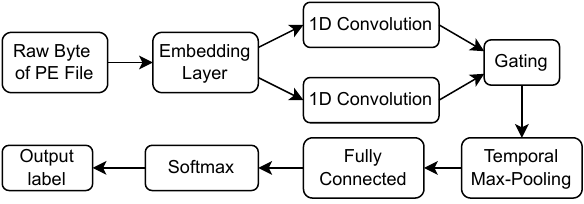}
    \caption{Architecture of MalConv\cite{raff2018malware} malware detector}
    \vspace{-5mm}
    \label{fig:malconv}
\end{figure}

Malware detection is a broad domain, however, we focus on widely used CNN-based end-to-end detection tailored for Windows PE Malware. Our target model of adversarial attack in this work is MalConv model~\cite{raff2018malware} , an academic standard to carry out adversarial attacks in the malware domain. MalConv learns to discriminate malware by taking the whole executable in the form of byte streams without undergoing any feature extraction process. Motivated to address the high amount of positional variation present in the executable files, Raff et al. came up with the MalConv convolution network architecture, as presented in Figure \ref{fig:malconv}. The combination of convolutional activation with max-pooling followed by fully connected layers allows the model to produce activation regardless of the location of the feature in the file. The input size to the model is fixed to $2^{21}$ bytes or 2MB. If the malware size is smaller, then it is padded to reach the threshold size, and in case of a larger file size, it's trimmed to take only the first ${2^{21}}$ bytes. The architecture maps the raw bytes to a fixed-length feature vector using an embedding layer before performing the convolution operation. 
\subsection{Adversarial Malware}

Since our work aims to increase the efficiency of adversarial evasion attacks using explainability, it's critical to understand the current state in adversarial malware research. After the early research ~\cite{anderson2018learning,kolosnjaji2018adversarial} that explored the possibility of adversarial attacks against malware detectors, numerous works have successfully crafted the adversarial attacks against malware detectors in problem space. The adversarial attacks in the malware domain can be distinguished from different bases depending on the perturbation generation algorithms used, target models, access to the target model, adversarial goals, adversary's capabilities, etc. However, within the domain of PE adversarial malware, two significant distinctions stand out: (a) the location for perturbation injection within PE malware, and (b) the preservation of file integrity. The location of perturbations during an adversarial evasion attack plays a critical role in preserving/breaking the file integrity, attack efficiency, and detection by malware detectors.

\begin{figure}[!t]
    \centering
    \includegraphics[scale=0.8]{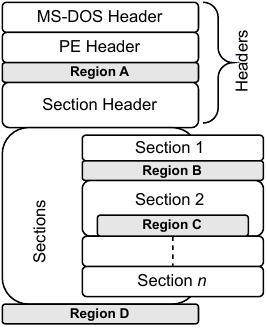}
    \caption{Adversarial attacks on different PE malware regions}
    \vspace{-5mm}
    \label{fig:attack_regions}
\end{figure}

The locations exploited by adversarial attackers to inject adversarial perturbation inside a PE file are represented in Figure \ref{fig:attack_regions} by \textit{Regions A-D}. Majority of the existing attacks in PE malware are append attacks (denoted by \textit{Region D} in Figure \ref{fig:attack_regions}), which appends the perturbation at the end of the file~\cite{kolosnjaji2018adversarial, kreuk2018deceiving, chen2019adversarial}. Despite the ability to not interfere with the integrity of malware files, perturbation at the end of the file has some challenges. For fixed-input malware detectors like MalConv~\cite{raff2018malware}, append attacks are ineffective~\cite{demetrio2019explaining, yuste2022optimization} for malware larger than 2MB. 
Attackers have leveraged \textit{slack space} (\textit{Region C} in Figure \ref{fig:attack_regions}), i.e. the unused spaces within a PE file not containing any meaningful data, to inject the perturbations~\cite{suciu2019exploring}. The adversarial injections in these spaces are more efficient than append attacks but are not guaranteed to be present in enough volume for every malware. Other approaches have found headers to be the most efficient location for adversarial perturbation~\cite{demetrio2019explaining, demetrio2021adversarial}, while some have found the injected code cave to edge ahead of the existing approaches~\cite{yuste2022optimization}. Recent work ~\cite{aryal2024intrasection} on intra-section code cave injection has shown higher success rates in attacking different regions of PE malware files. Analyzing existing literature, we can conjecture the trade-offs between the adversarial attack success, efficiency and the integrity of the malware detector. 

\begin{figure*}
    \centering
    \includegraphics[width=\textwidth]{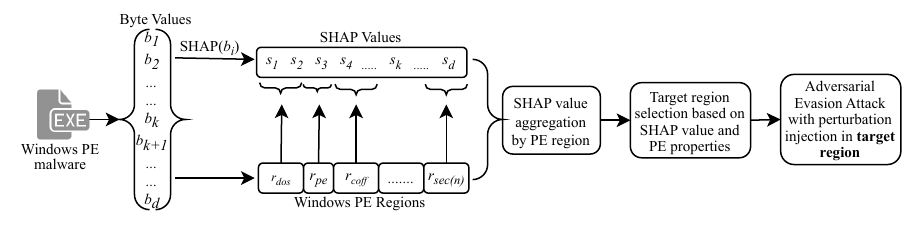}
    \vspace{-5mm}
    \caption{Proposed explainability guided  adversarial evasion attacks}
    \vspace{-5mm}
    \label{fig:approach_method}
\end{figure*}

\subsection{Explainability}
Explainability is a powerful tool for answering a \textit{black-box} machine learning system's questions of \textit{How?} and \textit{Why?}. It comprises of the processes and methods that help humans comprehend and trust the outputs of ML algorithms. As machine learning advances in making high-stakes decisions, clarifying the rationale behind the decisions is crucial rather than accepting them as a \textit{black-box} decision. It is easier to interpret the output of fundamental ML models, but the growth of complex/deep models has presented the trade-off between accuracy and interpretability of the model's output. Among numerous available explainable approaches, we are interested in exploring SHapley Additive exPlanation(SHAP) approach for model interpretation~\cite{lundberg2017unified}. SHAP assigns each feature an importance value by creating an \textit{explanation model} for a model's prediction. The simpler explanation model makes an interpretable approximation of the complex model whose decision needs to be explained. Explanation models work by simplifying the original input \(x\) to \(x^{\prime}\) with a mapping function \(x = h_{x}(x^{\prime})\). The additive feature attribution methods contain an explanation model that is a linear function of binary variables and is defined as:
\begin{equation}
g\left(z^{\prime}\right)=\phi_0+\sum_{i=1}^M \phi_i z_i^{\prime}
\end{equation}
where \(z^{\prime} \in\{0,1\}^{M}\), \(M\) is the total number of simplified input features and \(\phi_i \in \mathbb{R}\). 

SHAP values provide unique additive feature importance measures obtained from the Shapley values~\cite{shapley1953value} of a conditional expectation function of the original model. Our model of interest in this work is the Deep SHAP module of SHAP that combines Shapley values with DeepLIFT~\cite{shrikumar2017learning}. DeepLIFT uses a linear composition rule equivalent to linearizing the non-linear components of neural networks. With DeepLIFT being an additive feature attribution method satisfying local accuracy and missingness properties and Shapley values satisfying consistency properties, DeepLIFT leads to DeepSHAP through compositional approximation of SHAP values~\cite{lundberg2017unified}. Deep SHAP combines SHAP values for small components to the SHAP values of the entire network by recursively passing DeepLIFT's multipliers defined in terms of SHAP values. While dealing with simple network components like the linear or max-pooling layer with a single input, their SHAP values can be quickly computed, allowing a speedy approximation. Deep SHAP eliminates the need to simplify components by deriving a simplification from the SHAP values calculated for each component.

\section{Explainability guided adversarial evasion attacks}
\label{sec:exp-ae}
Our approach for explainability guided adversarial evasion attack on Windows PE malware detector, shown in Figure \ref{fig:approach_method}, can be divided into three major parts: (1) calculation of SHAP values for a given Windows PE malware byte stream, (2) utilizing the calculated SHAP values to strategize the selection of a target PE section for adversarial perturbation injection, and (3) crafting adversarial perturbation. Algorithm \ref{Alg:adve_attack} presents the flow of our approach from SHAP computation to adversarial perturbation creation. Our methodology is applicable to any machine learning-based malware detector, although our implementation specifically targets the MalConv model.

\begin{algorithm}[!t]
\caption{Explainability aided adversarial evasion attack}
\begin{algorithmic}[1]
    \State  \textbf{Input:} Malware PE file  with a byte streams $B$ = \{${b_{1}, b_{2},\cdots, b_{k}, \cdots, b_{d}}$\} where $d$ is total bytes which is 2000000 and $d-k$ is appended bytes; 
    \State \textbf{Output:} Modified malware file $B^{'}$ that can evade MalConv malware detector
    \State Compute SHAP values for each binary byte in $B$ using SHAP's DeepExplainer module , $SHAP(b_i)$
    \State Partition the malware binary stream $B$ as per Windows PE file structure: $Region(B)$ = \{${R_{dos}, R_{pe},\cdots, R_{sec(n)}}$\}, where each region corresponds to structural components like DOS Header, PE Signature, and so on. 
    \State Aggregate the SHAP values by PE region: 
    \[
    \operatorname{SHAP}_{\text {Agg }}(R)=\sum_{b_i \in R} \operatorname{SHAP}\left(b_i\right) 
    \]
    \State Based on aggregated SHAP values and the nature of the PE region towards modification(manually selected), target region $T$ for adversarial perturbation is chosen.
    \State At target region $T$, perturbation $\delta$ is generated using gradient descent until evasion or threshold condition, given $B^{'}$ = $B+\delta$ is a valid malware.
\end{algorithmic}
\label{Alg:adve_attack}
\end{algorithm}



\subsection{SHAP values for Windows PE malware byte stream}
The proposed approach starts with calculating SHAP values for the bytes sequence of a PE malware file as shown in Algorithm \ref{Alg:adve_attack}. The PE malware is presented as a sequence of bytes, $B$ = \{${b_{1}, b_{2},\cdots, b_{k}, \cdots, b_{d}}$\} with $d-k$ padding bytes as our target model, MalConv~\cite{raff2018malware} takes a fixed sized input of 2,000,000 bytes. We use \textit{DeepExplainer} (enhanced version of DeepLIFT) module of the SHAP library~\cite{lundberg2017unified} to approximate the conditional expectations of SHAP values using background samples. 
We modified the SHAP library to make it compatible with our target MalConv model for AE attack. The presence of an embedding layer that converts discrete bytes to continuous values stops the gradient from passing through this layer. To overcome this limitation, we computed SHAP values with respect to embedding output in place of an actual input byte stream. Additionally, the SHAP library was modified to handle multiple sigmoid layers present in MalConv. The \textit{DeepExplainer} module computes SHAP values for every corresponding binary byte of PE malware represented by the function $SHAP(b_i)$. These values of individual bytes provides insights into the contribution of each byte towards the detection of malware by MalConv. 

\subsection{Utilizing SHAP values to strategize the AE attack}
The computed SHAP values \{$ s_{1}, s_{2}, \cdots, {s_{d}}$\} corresponding to each input PE malware bytes can not be directly used without mapping to Windows PE structure since different regions/sections have different sensitivity to modification. The goal of injecting the perturbation is not only limited to evading a malware detector but also preserving the functionality of a malware file. The first step towards making meaningful strategy is by mapping SHAP values \{$ s_{1}, s_{2}, \cdots, {s_{d}}$\} of malware binary bytes to Windows PE file structure divided into different \textit{Region(B)}=  \{${R_{dos}, R_{pe},\cdots, R_{sec(n)}}$\}, presented in Step 3 of Algorithm \ref{Alg:adve_attack}. Each PE region is assigned an aggregated SHAP value of 
all the bytes in that region. The PE regions are placed in the descending order of aggregated SHAP value so that the region with maximum impact takes the first place while the one with minimum impact takes the last position in the list. At this point, the straightforward approach is to inject the perturbation in the region with the highest aggregated SHAP value, however, this approach is more complex and sensitive than simply injecting the perturbation based on the SHAP values order. 

Aggregated SHAP values provide a baseline order of the PE regions to choose a target for perturbation injection. However, in addition to SHAP values, other factors that affect the parsing of PE executable file format need to be adequately considered to preserve the file's integrity~\cite{nisi2021lost}. One of the first targets is the DOS header, since the only significance of this part of the header is backward compatibility and has no use in modern systems except for the magic number and the value at offset \textit{0x3c}. However, other than the DOS header, the header regions of Windows PE are very sensitive to modifications and can easily lead to breaking the malware file's executability. The other target region for perturbation is widely adopted append attacks, where the perturbation is appended at the end of the file without impacting the malware's behaviour. Recent research has explored the possibility of injecting the perturbation in different sections of Windows PE malware without impacting the file. Work by Aryal et al.\cite{aryal2024intrasection} have already shown successful attacks while injecting perturbation in the \texttt{.text}, \texttt{.data}, \texttt{.rdata} and other sections of PE file. Therefore, we also use the SHAP values to inject the perturbation in these sections. We chose the section with the highest SHAP value, i.e., the highest contribution, as a target section for an efficient adversarial evasion attack.

We do not limit the utilization of SHAP value just for choosing the target section but to choose the most impactful subsection inside the target section. As an example, the average size of \texttt{.text} section on our PE malware dataset is found to be 97,000 bytes, which is 10\%  of the total malware size but a larger perturbation space. To devise a suitable target location for perturbation within an ample section space, we divided each PE section into smaller fixed-size subsections and aggregate the shape values for each corresponding subsection. The smaller subsections with larger SHAP values is chosen as a target region for injecting the perturbation. This approach gives us finer control over the perturbation location even within the same section of the PE malware, enabling more efficient adversarial evasion attacks on Windows PE malware.

\begin{figure}
    \centering
    \includegraphics[width=0.7\linewidth]{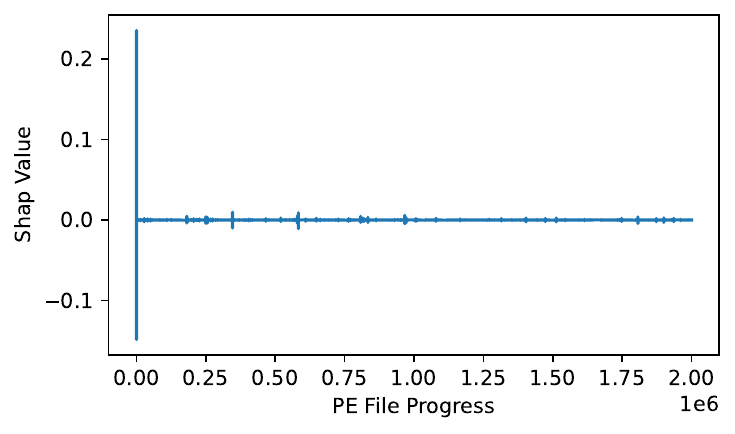}
    
    \tiny\textit{{\textbf{Malware SHA-256 Hash}:0DC8473AAF3522E278EA057C79764E5E7F2EC6DE8E29D95D89A410E224278612}}
    \caption{Local SHAP values for an instance of Windows PE Malware}
    \label{fig:local_shap_full}
\end{figure}

\begin{figure}
    \centering
    \includegraphics[width=0.79\linewidth]{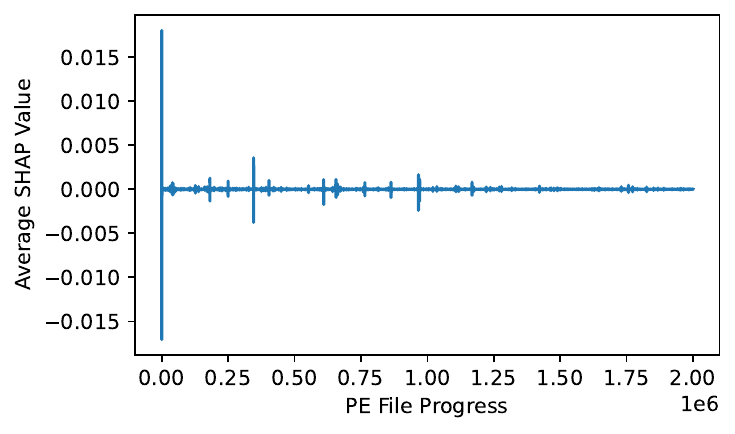}
    \caption{Global average SHAP values for 500 Windows PE malware}
    \vspace{-5mm}
    \label{fig:global_shap_full}
\end{figure}

\subsection{Crafting adversarial perturbation}
At this stage, we execute the attack process on the target region provided by the previous stage. However, due to its nature, each PE region has its own requirement to perform the functionality-preserving attack. Only specific bytes are modified for attacks in the DOS header; for the append attack, the length of perturbations is provided. To inject the perturbations inside the PE sections, the code caves and code loaders are injected\cite{aryal2024intrasection} to the PE file.
The target regions can now hold any perturbation without altering the integrity of the PE malware file. To generate adversarial perturbation, we adopted Kolosnjaji et al.'s\cite{kolosnjaji2018adversarial} approach of gradient descent. The approach first computes the embedding representation of input malware bytes(${B}$) as $\mathbf{Z} \leftarrow \phi(\boldsymbol{b})$. The non-differentiability of our target MalConv model is resolved by computing the negative gradient of the loss function with respect to embedding representation($\mathbf{Z}$) given by, $\boldsymbol{w}_i=-\nabla_\phi\left(b_i\right) \in \mathbb{R}^8$. The approach will select the closest byte in the embedding space that will maximally increase the probability of misclassifying malware by the malware detector. 

\begin{figure*}
    \centering
    \includegraphics[width = 0.79\textwidth]{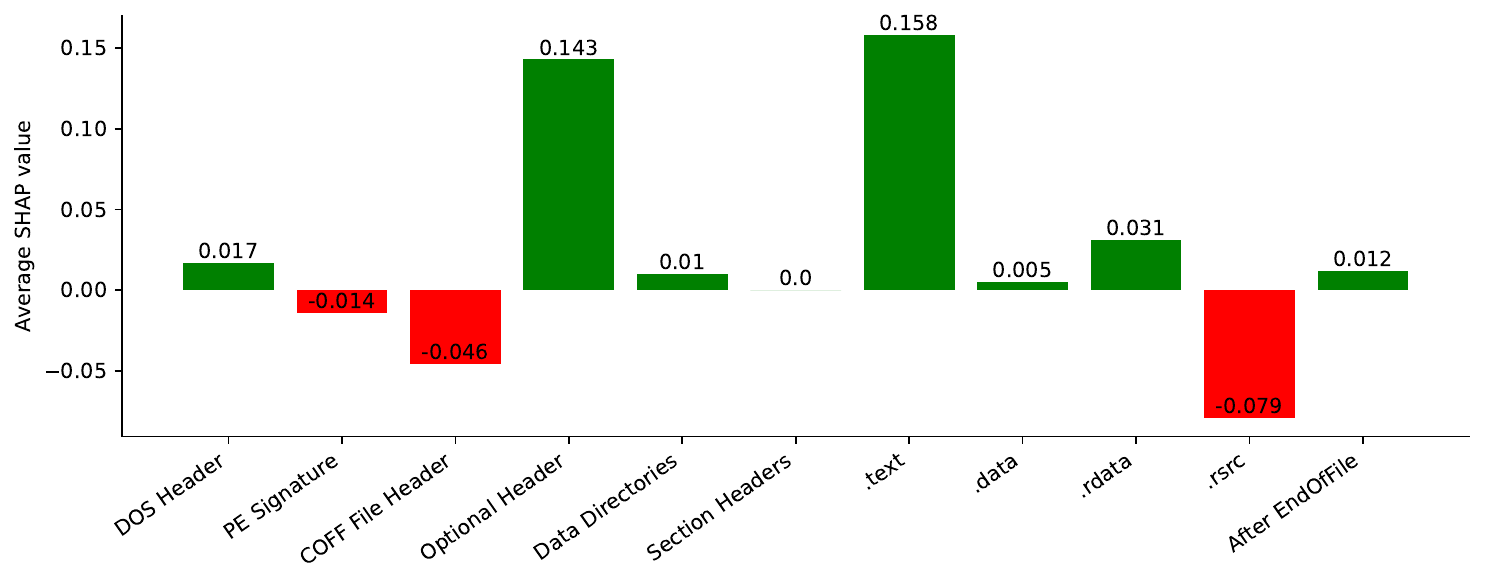}
    
    \tiny\textit{{\textbf{Malware SHA-256 Hash}:0DC8473AAF3522E278EA057C79764E5E7F2EC6DE8E29D95D89A410E224278612}}
    \vspace{-2mm}
    \caption{Local Interpretation SHAP values mapped to different Windows PE malware regions}
    \label{fig:local_shap_section}
\end{figure*}

\begin{figure*}
    \centering
    \includegraphics[width = 0.77\textwidth]{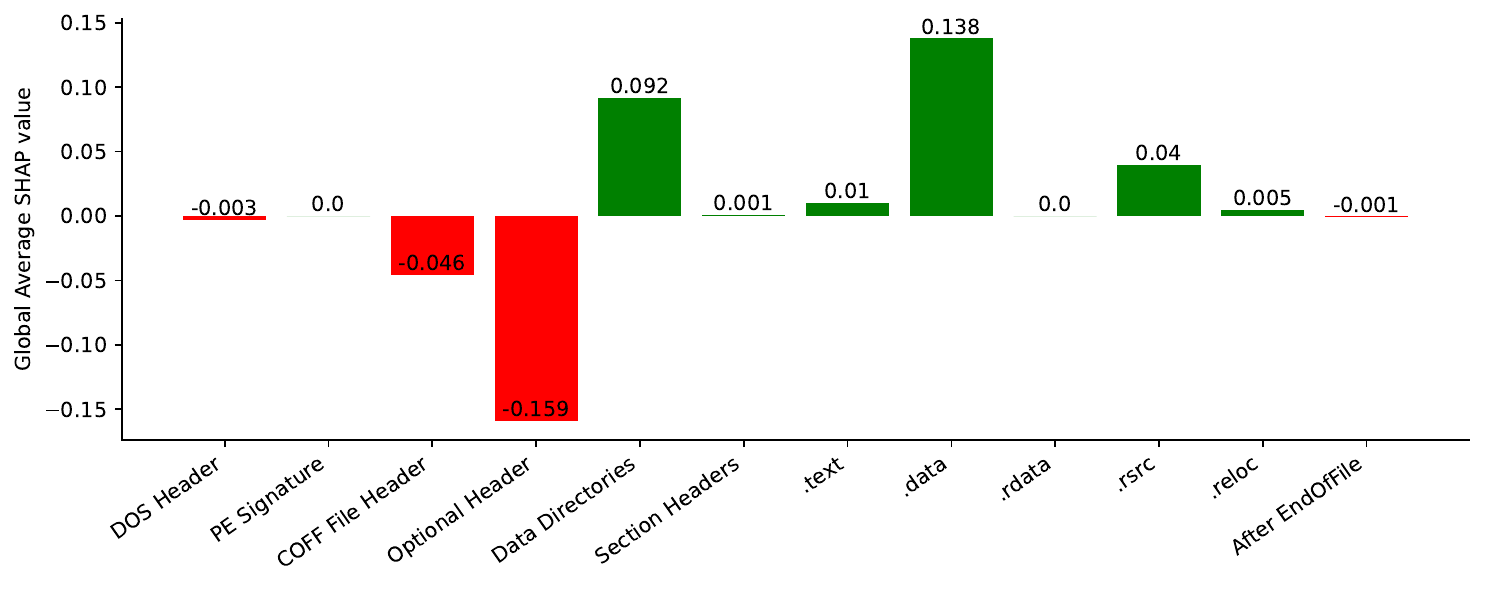}
    \vspace{-2mm}
    \caption{Global Interpretation Average SHAP values mapped to different Windows PE malware regions}
    \vspace{-5mm}
    \label{fig:global_shap_section}
\end{figure*}

\section{Experiments and Evaluation}
\label{sec:experiment}
Our experimental setup consists of 6000 Windows PE malware from VirusTotal~\cite{VirusTot69:online}, out of which 5000 malware are used to train the MalConv model and the rest 1000 for validation and evaluation purposes. We verified the malware files to determine if they were encrypted by calculating the entropy of \texttt{.text} section. We found only 2\% of malware to have entropy greater than 7.2, indicating that they might be encrypted. We reached end-to-end Windows PE malware detection accuracy of 96\% with the MalConv model, which is used as a target model to compute SHAP values leading to adversarial evasion attacks. We used 100 malware as background data for SHAP's \textit{DeepExplainer} module and computed SHAP values for 500 malware. We used \textit{pefile} ~\cite{pefileonline}  python library to gain insights into PE malware file structure. The following \textbf{research question} guided our evaluation.
\begin{itemize}
    \item \textbf{RQ1 - Local vs. Global explanation:} Can a global explanation profile the SHAP values of a local PE malware sample and help create an adversarial evasion against MalConv?
    \item \textbf{RQ2 - SHAP value vs. Evasion rate:} How do the SHAP values of different PE regions relate to the adversarial evasion rate?
    \item \textbf{RQ3 - Granular SHAP analysis within the same section:} Do the adversarial injection targets in different locations within the same section of PE malware impact attack results?
\end{itemize}

\begin{table*}[]
\centering
\caption{Aggregated SHAP value vs. Evasion rate on crafting adversarial sample with different PE target region}
\vspace{-2mm}
\begin{tabular}{|l|ll|ll|ll|ll|}
\hline
\multicolumn{1}{|c|}{\multirow{2}{*}{\textbf{Target Region}}} &
  \multicolumn{2}{c|}{\textbf{Average Size of}} &
  \multicolumn{2}{c|}{\textbf{Aggregated SHAP value}} &
  \multicolumn{2}{c|}{\textbf{iteration = 20}} &
  \multicolumn{2}{c|}{\textbf{iteration = 40}} \\ \cline{2-9} 
\multicolumn{1}{|c|}{} &
  \multicolumn{1}{c|}{\textbf{Section}} &
  \multicolumn{1}{c|}{\textbf{Perturbation}} &
  \multicolumn{1}{c|}{\textbf{Mean}} &
  \multicolumn{1}{c|}{\textbf{Abs. Mean}} &
  \multicolumn{1}{c|}{\textbf{Evasion Rate}} &
  \multicolumn{1}{c|}{\textbf{Confidence}} &
  \multicolumn{1}{c|}{\textbf{Evasion Rate}} &
  \multicolumn{1}{c|}{\textbf{Confidence}} \\ \hline
DOS Header        & \multicolumn{1}{l|}{64}    & 64   & \multicolumn{1}{l|}{-0.0035} & 0.0251 & \multicolumn{1}{l|}{9.23\%}  & 0.84 & \multicolumn{1}{l|}{8.06\%}  & 0.85 \\ \hline
PE Signature      & \multicolumn{1}{l|}{4}     & 4    & \multicolumn{1}{l|}{0}       & 0.0291 & \multicolumn{1}{l|}{2.75\%}  & 0.96 & \multicolumn{1}{l|}{2.75\%}  & 0.96 \\ \hline
COFF File Header  & \multicolumn{1}{l|}{20}    & 20   & \multicolumn{1}{l|}{-0.0474} & 0.2909 & \multicolumn{1}{l|}{3.93\%}  & 0.93 & \multicolumn{1}{l|}{3.93\%}  & 0.93 \\ \hline
Optional Header   & \multicolumn{1}{l|}{223}   & 223  & \multicolumn{1}{l|}{-0.1636} & 1.5982 & \multicolumn{1}{l|}{75.05\%} & 0.47 & \multicolumn{1}{l|}{75.64\%} & 0.46 \\ \hline
Data Directories  & \multicolumn{1}{l|}{127}   & 127  & \multicolumn{1}{l|}{0.0903}  & 0.4953 & \multicolumn{1}{l|}{19.45\%} & 0.84 & \multicolumn{1}{l|}{18.27\%} & 0.84 \\ \hline
Section Headers   & \multicolumn{1}{l|}{127}   & 127  & \multicolumn{1}{l|}{0.0013}  & 0.0113 & \multicolumn{1}{l|}{2.75\%}  & 0.96 & \multicolumn{1}{l|}{2.75\%}  & 0.96 \\ \hline
.text             & \multicolumn{1}{l|}{96992} & 3822 & \multicolumn{1}{l|}{0.0004}  & 0.1008 & \multicolumn{1}{l|}{22.92\%} & 0.82 & \multicolumn{1}{l|}{21.13\%} & 0.84 \\ \hline
.data             & \multicolumn{1}{l|}{41110} & 3550 & \multicolumn{1}{l|}{0.0115}  & 0.21   & \multicolumn{1}{l|}{16.67\%} & 0.86 & \multicolumn{1}{l|}{15.76\%} & 0.87 \\ \hline
.rdata            & \multicolumn{1}{l|}{18537} & 3172 & \multicolumn{1}{l|}{-0.0003} & 0.1473 & \multicolumn{1}{l|}{41.37\%} & 0.65 & \multicolumn{1}{l|}{42.81\%} & 0.64 \\ \hline
.rsrc             & \multicolumn{1}{l|}{63611} & 3607 & \multicolumn{1}{l|}{0.0023}  & 0.0704 & \multicolumn{1}{l|}{30.27\%} & 0.75 & \multicolumn{1}{l|}{28.65\%} & 0.76 \\ \hline
.reloc            & \multicolumn{1}{l|}{13006} & 2627 & \multicolumn{1}{l|}{0.001}   & 0.1054 & \multicolumn{1}{l|}{37.42\%} & 0.67 & \multicolumn{1}{l|}{37.42\%} & 0.68 \\ \hline
After End Of File & \multicolumn{1}{c|}{-}     & 3863 & \multicolumn{1}{l|}{-0.0002} & 0.0404 & \multicolumn{1}{l|}{7.07\%}  & 0.93 & \multicolumn{1}{l|}{7.66\%}  & 0.93 \\ \hline
\end{tabular}
\vspace{-5mm}
\label{Table:SHAPvsEvasion}
\end{table*}

\subsection{Local vs. Global explanation}
As discussed, we calculated the local SHAP for a single instance of PE malware and global SHAP averaged over 500 malware as shown in Figures \ref{fig:local_shap_full} and \ref{fig:global_shap_full} respectively.
These figures show the SHAP value for each byte of PE malware binary for its entire 2000000 input bytes in the order of their position in the file. The local and global explanations exhibit a similar pattern of SHAP values as they progress through the PE malware bytes. SHAP values for both the local and global SHAP plots in  Figures \ref{fig:local_shap_full} and  \ref{fig:global_shap_full} show significantly higher values at initial regions and attenuate as they progress through the file. However, we can see some spikes in SHAP value across the byte sequence in regions in later part as well. The SHAP value spikes in different locations of the file of various lengths vary in each PE malware file, evident from the non-identical sized spike in Global SHAP (Figure \ref{fig:global_shap_full}) compared to local SHAP (Figure \ref{fig:local_shap_full}). These results show that the distinct regions of the PE malware binary have different contributions towards being detected, and the contribution of areas also varies as per the individual malware file.  

To further assess the distribution of SHAP values in byte streams of Windows PE malware, we plotted the aggregated SHAP value across different regions of the Windows PE malware structure. Figure \ref{fig:local_shap_section} shows aggregated SHAP values across different regions of a single local instance of Windows PE malware. In contrast, Figure \ref{fig:global_shap_section} shows the average aggregated SHAP of different PE malware regions across malware samples. The green bars in figures represent positive contribution towards malware detection, while the red bars represent the negative contribution. In Figure \ref{fig:local_shap_section}, it can noticed that the \textit{DOS header}, \textit{Optional Header}, \texttt{.text} , \texttt{.rdata},  \texttt{.data} sections and the end of file contents contribute towards the detection and the rest towards making malware detected as `benign'. In both plots, we should not forget that the aggregated SHAP value results from different-sized PE regions. The aggregated SHAP contribution of the DOS header is the result of 64 bytes while the SHAP value of \texttt{.text} section is from 97,000 bytes on average. Our results in the next section explain the efficiency of attacks in different regions.

We made some contrasting observations in the Local and Global explanation plots of SHAP values in Figure \ref{fig:local_shap_section} and \ref{fig:global_shap_section}. The aggregated SHAP values of the PE malware section from the local instance are inconsistent with the global averaged SHAP values. As an example, \texttt{.rsrc} section has a positive SHAP value in the local instance (Figure \ref{fig:local_shap_section}) while having a negative value in global averaged SHAP for entire malware. This behaviour is due to variations in the size of malware and its sections. A particular PE region could be at different locations in individual PE files. 

\begin{tcolorbox}[colback=white, boxrule=1pt, arc=0pt,top = 0pt, bottom = 0pt, left=0pt, right=0pt]
    \textbf{Answer to RQ1:} \textit{As observed, each PE malware file differs in the size and location of its regions; a global averaged explanation for PE regions can not be created against the CNN-based malware detector, MalConv. The contrasting observation of SHAP values for individual regions in Figure \ref{fig:local_shap_section} and \ref{fig:global_shap_section} demonstrates that each malware needs a local explanation to profile the SHAP values of its PE regions precisely.}
\end{tcolorbox}

\subsection{SHAP value vs. Evasion rate}
\begin{figure}
    \centering
    \includegraphics[width=\linewidth]{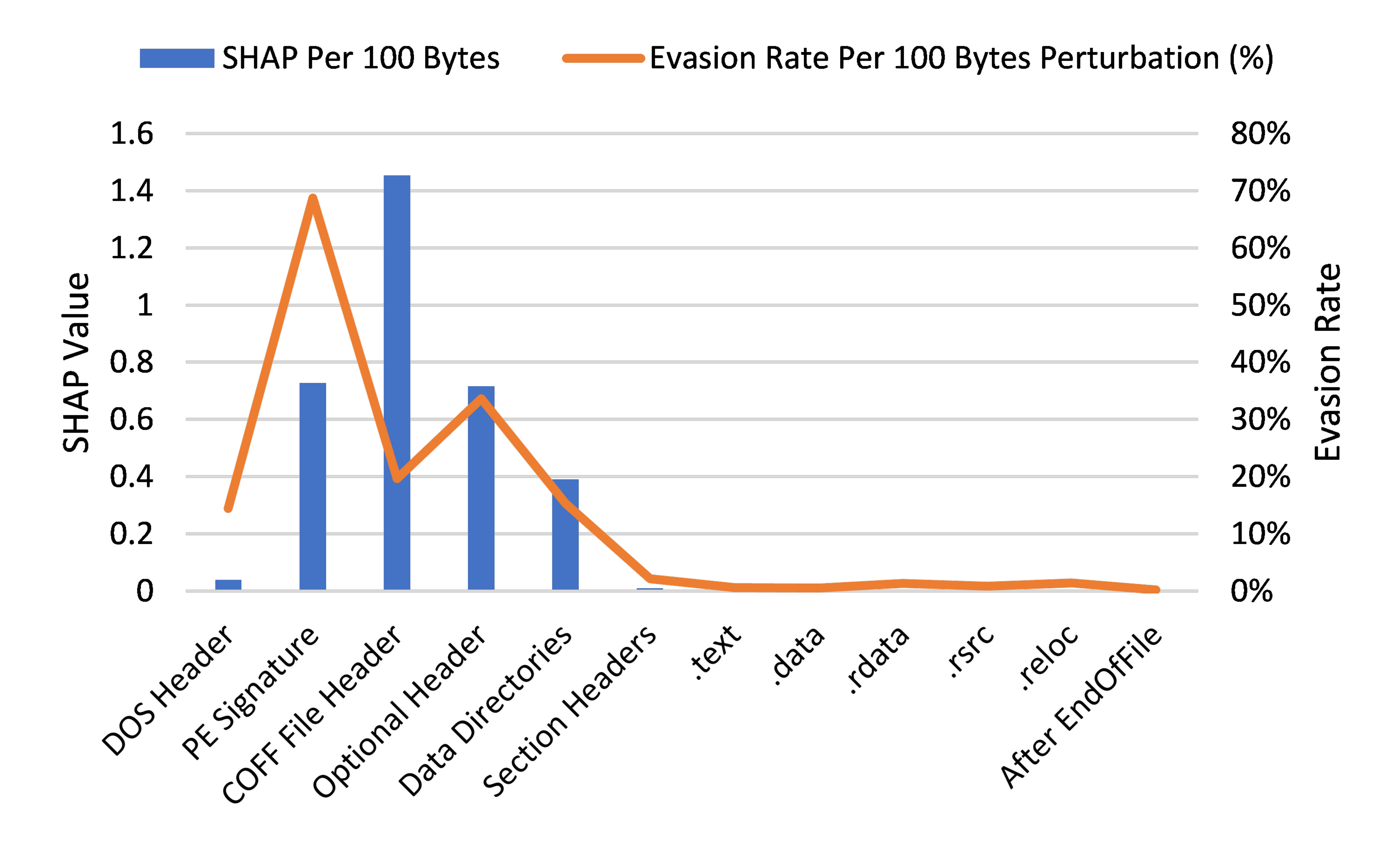}
    \caption{Plotting SHAP values vs. Evasion rate scaled for 100 bytes for different PE regions }
    \vspace{-4mm}
    \label{fig:SHAPvsEvasion}
\end{figure}

\begin{figure*}
    \centering
    \includegraphics[width = \textwidth]{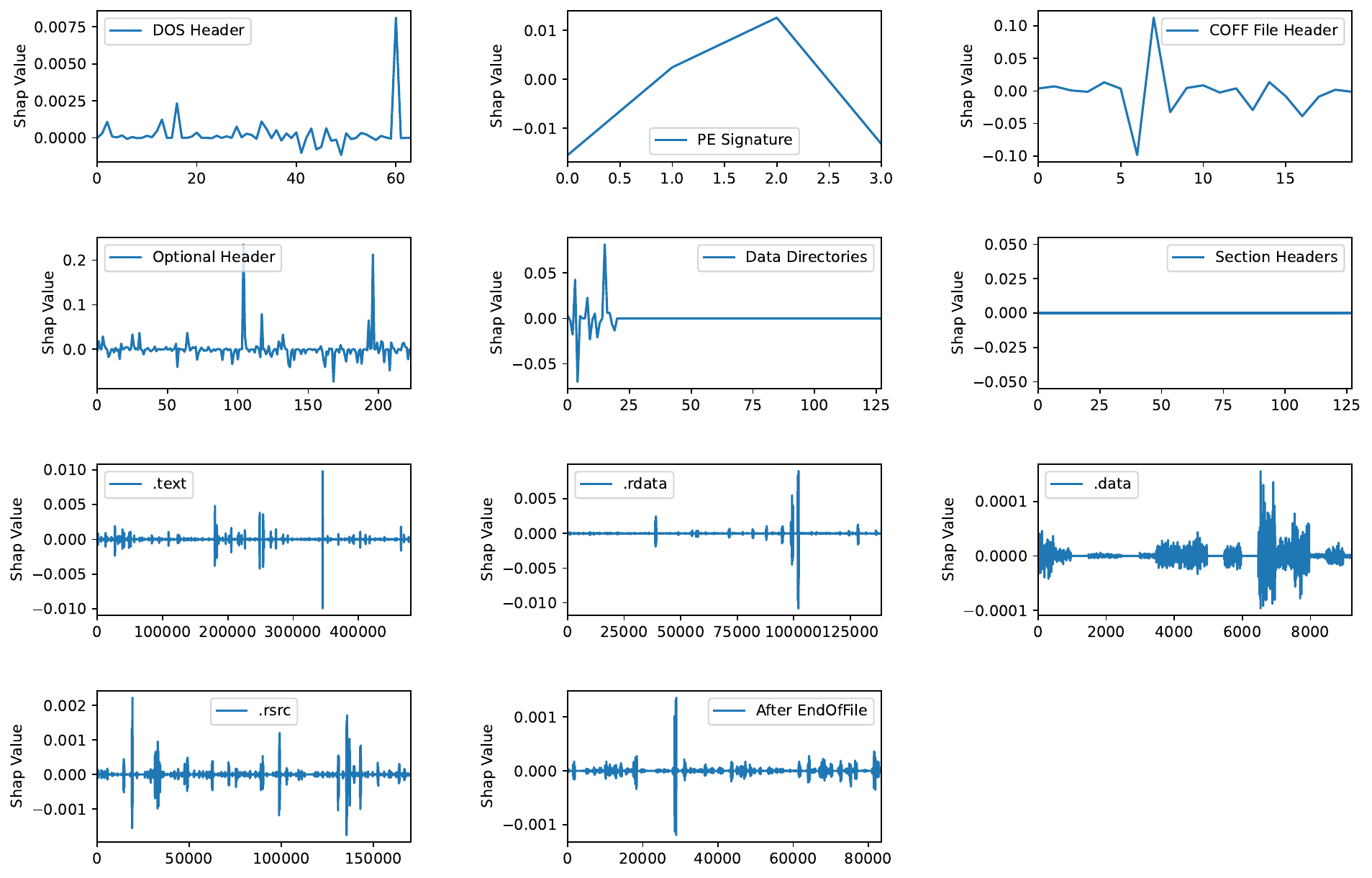}
    \footnotesize\textit{{\textbf{Malware SHA-256 Hash}:0DC8473AAF3522E278EA057C79764E5E7F2EC6DE8E29D95D89A410E224278612}}
    \caption{SHAP values within each Windows PE malware region across its length}
    \vspace{-3mm}
    \label{Fig:SHAP_Section_Granular}
\end{figure*}
Here, we inject perturbation in different regions of Windows PE malware, aiming to examine the correlation between the corresponding SHAP values and the effectiveness of the adversarial evasion attack against the MalConv. While the functionality-preserving attack focuses solely on injecting the perturbation into the \textit{DOS Header}, \texttt{.text}, \texttt{.data}, \texttt{.rdata}, \texttt{.rsrc} sections, and at the \textit{end of the PE malware file}, in order to extract the precise relation between SHAP values and evasion rate, we conducted adversarial injections across all the feasible regions of PE malware. We understand that injecting the perturbation into the \textit{PE signature, COFF file header, Optional header, Data directories and Section headers} compromises the integrity of the file in adversarial malware samples. Since we have to analyze relatively small header regions of the PE file, we limit the perturbation size to 4096 bytes for each PE region across the malware samples.

Table \ref{Table:SHAPvsEvasion} shows results for evasion rates and average confidence of MalConv on PE malware samples after injecting the average size perturbation. We compared by running gradient descent for up to 40 iterations, and Table \ref{Table:SHAPvsEvasion} shows the result for 20 and 40 iterations. However, it is observed that there is no significant change in either the evasion rate or the average confidence while increasing the iterations from 20 to 40. We made another intriguing observation while linking SHAP values with the evasion rate. It is found that the perturbations in the PE regions with negative SHAP values can evade the malware detector at a similar rate to those perturbations in PE regions with positive SHAP values. By calculating the absolute mean of SHAP values for each region in the PE malware structure, we were able to establish a more discrete relationship with the evasion rate. Since the gradient descent optimizes each perturbation equally, the distinction between positive and negative impact (SHAP values) became irrelevant. Instead, only the magnitude of contribution (SHAP value) to the decision-making process is relevant. From the Table \ref{Table:SHAPvsEvasion}, it is evident that the absolute mean SHAP values align more closely with the evasion rate as compared to the mean SHAP values. 

\begin{table*}[]
\centering
\caption{Results from adversarial injection attack on the granular first subsection, subsection with least SHAP and the subsection with highest SHAP}
\begin{tabular}{|l|lll|lll|lll|}
\hline
\multicolumn{1}{|c|}{\multirow{2}{*}{\textbf{\begin{tabular}[c]{@{}c@{}}Target\\ Section\end{tabular}}}} &
  \multicolumn{3}{c|}{\textbf{First subsection}} &
  \multicolumn{3}{c|}{\textbf{Lowest Agg SHAP subsection}} &
  \multicolumn{3}{c|}{\textbf{Highest Agg SHAP subsection}} \\ \cline{2-10} 
\multicolumn{1}{|c|}{} &
  \multicolumn{1}{c|}{\textbf{abs(SHAP)}} &
  \multicolumn{1}{c|}{\textbf{Evasion Rate}} &
  \multicolumn{1}{c|}{\textbf{Avg. Conf.}} &
  \multicolumn{1}{c|}{\textbf{abs(SHAP)}} &
  \multicolumn{1}{c|}{\textbf{Evasion Rate}} &
  \multicolumn{1}{c|}{\textbf{Avg. Conf.}} &
  \multicolumn{1}{c|}{\textbf{abs(SHAP)}} &
  \multicolumn{1}{c|}{\textbf{Evasion Rate}} &
  \multicolumn{1}{c|}{\textbf{Avg. Conf.}} \\ \hline
  .text &
  \multicolumn{1}{l|}{0.181} &
  \multicolumn{1}{l|}{27.45\%} &
  0.7883 &
  \multicolumn{1}{l|}{0.0258} &
  \multicolumn{1}{l|}{17.65\%} &
  0.8518 &
  \multicolumn{1}{l|}{0.5251} &
  \multicolumn{1}{l|}{57.35\%} &
  0.5924 \\ \hline
.data &
  \multicolumn{1}{l|}{0.2275} &
  \multicolumn{1}{l|}{31.25\%} &
  0.7525 &
  \multicolumn{1}{l|}{0.0459} &
  \multicolumn{1}{l|}{10.16\%} &
  0.897 &
  \multicolumn{1}{l|}{0.9983} &
  \multicolumn{1}{l|}{35.94\%} &
  0.7241 \\ \hline
.rdata &
  \multicolumn{1}{l|}{0.3652} &
  \multicolumn{1}{l|}{38.78\%} &
  0.6889 &
  \multicolumn{1}{l|}{0.031} &
  \multicolumn{1}{l|}{12.24\%} &
  0.8748 &
  \multicolumn{1}{l|}{0.7893} &
  \multicolumn{1}{l|}{48.98\%} &
  0.5983 \\ \hline
.rsrc &
  \multicolumn{1}{l|}{0.0953} &
  \multicolumn{1}{l|}{23.14\%} &
  0.82 &
  \multicolumn{1}{l|}{0.0182} &
  \multicolumn{1}{l|}{18.78\%} &
  0.8329 &
  \multicolumn{1}{l|}{0.4263} &
  \multicolumn{1}{l|}{37.99\%} &
  0.6923 \\ \hline

\end{tabular}
\vspace{-4mm}
\label{Table:granular_result}
\end{table*}

As shown in Table \ref{Table:SHAPvsEvasion}, we observed the evasion rate as high as 75.64\% on injecting only 223 byte average perturbation. Specifically, the perturbations injected in \texttt{.text}, \texttt{.data}, \texttt{.rdata}, \texttt{.rsrc} and \texttt{.reloc} sections also resulted a signification evasion rates of approximately 22\%, 16\%, 42\%, 30\% and 37\%, respectively. On the other hand, we can clearly see the evasion rate is negligible for adversarial injections in \textit{PE Signature}, \textit{COFF File header}, \textit{Section Header} and after \textit{end of the file} region. Since different regions mentioned in Table \ref{Table:SHAPvsEvasion} vary largely in terms of perturbation size and SHAP value, it does not provide a clear picture of attack efficiency for different PE malware regions. To better observe the attack efficiency and establish distinct relationship between SHAP values and evasion rate, we plotted a graph for aggregated absolute SHAP value versus the evasion rate, with both scaled to 100 bytes perturbation, as shown in Figure \ref{fig:SHAPvsEvasion}. The figure clearly shows a significantly higher efficiency of evasion in PE file header regions than in the rest of the PE body. The observation demonstrates a positive correlation between absolute SHAP values and the evasion rate.

\begin{tcolorbox}[colback=white, boxrule=1pt, arc=0pt,top = 0pt, bottom = 0pt, left=0pt, right=0pt]
    \textbf{Answer to RQ2:} \textit{The absolute SHAP value of the PE malware region exhibits approximately a positive correlation with the evasion rate as observed in Figure \ref{fig:SHAPvsEvasion}. Specifically, the higher the absolute SHAP for a PE region, the higher the evasion rate upon injecting adversarial perturbation.}
\end{tcolorbox}
\vspace{-4mm}

\subsection{Breaking SHAP aggregation within a PE malware section}
The disparity observed in the SHAP values among different regions within the PE malware file encouraged us to evaluate the variation of SHAP values even within the same PE malware region. The diversity shown in the subplots in Figure \ref{Fig:SHAP_Section_Granular}, demonstrates the SHAP values change within the same region to a large extent. We can easily see the places with significantly higher SHAP values than the other places in each subplot of Figure \ref{Fig:SHAP_Section_Granular}. This part of the evaluation is to quantify the difference in the success of AE attacks while choosing different perturbation locations within the same PE section based on the SHAP value. 

We subdivide the larger PE sections (over 10,000 bytes) of PE malware into smaller subsections, each of 4096 bytes. Now, we inject perturbation into three different subsections individually in each section of a PE malware file: (1) the first subsection, (2) the subsection with the lowest aggregated absolute SHAP, and (3) the subsection with the highest aggregated absolute SHAP. Table \ref{Table:granular_result} presents the results on injecting perturbation to different subsections inside the \texttt{.text}, \texttt{.data}, \texttt{.rdata} and \texttt{.rsrc} section of PE file. It is noted that the evasion rate on injecting perturbation in the first 4096 bytes of \texttt{.text} is 27.45\%, which dips to 17.65\% on using the subsection with the lowest SHAP while increasing up to 57.35\% on using the subsection with the highest SHAP. The difference in results varies from 10.16\% to 35.94\% in \texttt{.data}, from 12.24\% to 48.98\% in \texttt{.rdata} and from 18.78\% to 37.99\% in \texttt{.rsrc} sections while taking the lowest and highest SHAP aggregating subsections as a target for adversarial perturbation. A similar difference is also observed in the average confidence of the PE malware file while injecting perturbation in different subsections based on the SHAP value. 

\begin{tcolorbox}[colback=white, boxrule=1pt, arc=0pt,top = 0pt, bottom = 0pt, left=0pt, right=0pt]
    \textbf{Answer to RQ3:} \textit{The selection of subsections within the same sections of Windows PE malware for perturbation injections makes a significant difference in the adversarial evasion success. It is observed that the evasion rate taking a massive leap while choosing the subsection with higher aggregated SHAP values than those with lower aggregate SHAP values.}
\end{tcolorbox}

\section{Conclusion}
\label{sec:summary}
This study focuses on the use of explainability of machine learning-based malware detectors in crafting evasive adversarial PE malware samples. Leveraging a CNN-based MalConv malware detector, we demonstrated how the attributions provided by explainable algorithms, such as the SHAP, to various regions of Windows PE malware files, could be exploited to devise strategies for effectively injecting perturbations. Through our analysis, we extracted the relationship between the aggregated SHAP values and success of AE attacks. By discovering the most impactful regions within the PE file for perturbation injection, our findings highlight the significance of explainability in augmenting the sophistication and efficacy of adversarial evasion techniques for Windows PE malware detection. To gain a more fine-grained analysis, we divided PE sections into smaller subsections to inject perturbations based on the preference decided by SHAP value. In future research, we aim to broaden the scope of explainability by exploring its application in conducting black box attacks, leveraging the concept of transferability. Moreover, we will explore additional efficient explainable tools to enhance adversarial attacks within the adversarial malware domain.



\section*{Acknowledgment}
This work is partially supported by the NSF grants 2230609 and 2230610.

\bibliographystyle{IEEEtran}
\bibliography{biblio}
\end{document}